\documentclass[%
 nofootinbib,
 amsmath,amssymb,
 aps,
 prd,
]{revtex4-2}

\usepackage{graphicx}
\usepackage{dcolumn}
\usepackage{slashed}
\usepackage{bm}
\usepackage{hyperref}
\usepackage[mathlines]{lineno}
\usepackage{makecell} 
\usepackage{multirow}
\usepackage{aas_macros}

\newcommand{\Lamd}{\ensuremath{\Lambda_{\rm D}}}

\newcommand{\zkd}{\ensuremath{z_{\rm kd}}}

\newcommand{\CLASS}{\textsc{Class}}

\begin{document}

\title{Dark Radiation Sticks Together: Dark QCD and the Hubble Tension}

\author{Matthew R.~Buckley}
\email{mbuckley@physics.rutgers.edu}
\affiliation{NHETC, Department of Physics and Astronomy, Rutgers, Piscataway, NJ 08854, USA}

\author{Nicolas Fernandez}
\email{nico.fer@rutgers.edu}
\affiliation{NHETC, Department of Physics and Astronomy, Rutgers, Piscataway, NJ 08854, USA}

\author{Eric Putney}
\email{eputney@physics.rutgers.edu}
\affiliation{NHETC, Department of Physics and Astronomy, Rutgers, Piscataway, NJ 08854, USA}

\date{\today}

\begin{abstract}
We introduce a model of dark matter charged under a new confining gauge group akin to the Standard Model's quantum chromodynamics (QCD). The fermionic sector contains both heavy and light quark-like fermions charged under the dark QCD. As the Universe cools, the gauge force confines and dark radiation transitions from free particles to low-mass dark pions. This form of dark radiation behaves as an imperfect fluid with nonzero viscosity. If the confinement occurs at the MeV scale, the model can alleviate the tension between early- and late-time measurements of the Hubble parameter, with a Bayesian posterior of $H_0 = 71.60\pm 0.63$~km/s/Mpc and a $3\sigma$ residual tension across {\em Planck}, ACT, BBN, Pantheon+, DESI DR2, and SH0ES. In the late Universe, the heavy dark matter can evade existing constraints on dark matter self-interactions while containing a rich phenomenology that may be accessible in future observations.
\end{abstract}

\maketitle

\section{Introduction \label{sec:into}}

Modern observations of the Universe are remarkably well-fit by the Cosmological Constant-Cold Dark Matter ($\Lambda$CDM) model. However, in recent years a number of small (but increasingly significant) tensions have arisen in precision measurements. The most notable and long-standing of these is the Hubble Tension (see \cite{CosmoVerseNetwork:2025alb} for a recent review), the $\sim 7\sigma$ difference between measurements of the Hubble parameter $H_0$ as measured by Type-1a supernovae ($H_0 = 73.04\pm 1.04$~km/s/Mpc \cite{Riess:2021jrx} from SH0ES alone and $73.50\pm 0.81$~km/s/Mpc \cite{H0DN:2025lyy} when combining datasets) and the value of $H_0$ inferred from measurements of the Cosmic Microwave Background (CMB) ($H_0 = 67.36\pm 0.54$~km/s/Mpc \cite{Planck:2018vyg} measured by {\em Planck}  and  $67.19\pm 0.38$~km/s/Mpc \cite{SPT-3G:2025bzu} from SPT).

Barring unidentified systematics in the observation and analysis pipeline, this tension appears to requires new physics prior to recombination and release of the CMB photons \cite{Knox:2019rjx}. Many such solutions have been suggested (see \cite{Knox:2019rjx,Schoneberg:2021qvd,CosmoVerseNetwork:2025alb,Schoneberg:2026eys,Schoneberg:2026vaf} for reviews), some of which can reduce the tension to $2-3\sigma$. Of particular relevance for the present work are models that increase the number of effective relativistic degrees of freedom $N_{\rm eff}$ \cite{Lesgourgues:2015wza,Bernal:2016gxb,Aloni:2021eaq,Escudero:2021rfi,Joseph:2022jsf,Aloni:2023tff,Gariazzo:2023hch,Schoneberg:2023rnx,Sobotka:2023bzr,Bagherian:2024obh,Allali:2024cji,Allali:2024cji}. This decreases the sound horizon $r_s$ and thus allows the Planck measurements of the CMB to accommodate larger values of $H_0$. However, a global increase of $N_{\rm eff}$ alone is disfavored by the high-$\ell$ CMB multipoles, and a recent meta-analysis of solutions to the Hubble tension \cite{Schoneberg:2026eys,Schoneberg:2026vaf} find that existing models that introduce new relativistic degrees of freedom can only improve the Hubble tension to the $\sim 4\sigma$ level.

In this work, we propose a new model containing extra relativistic degrees of freedom with unique features. We propose a confining quantum chromodynamics-like (QCD) dark force acting on dark matter and dark radiation. Such a model has novel phenomenology for the dark sector both during decoupling and in the present day; the behavior in the early Universe improves the agreement with cosmological data and eases the Hubble tension. In contrast to most existing models of dark QCD \cite{Nussinov:1985xr,Chivukula:1989qb,Barr:1990ca,Bagnasco:1993st,Banks:2005hc,Gudnason:2006ug,Hamaguchi:2007rb,Mardon:2009gw,Kribs:2009fy,Bai:2010qg,Frigerio:2012uc,Buckley:2012ky,Barnard:2014tla,Asadi:2026mip}, the confinement scale $\Lambda_{\rm D}$ is ${\cal O}({\rm MeV})$. As in the Standard Model, we assume the dark sector has both very heavy $m_Q \gg \Lambda_{\rm D}$ and very light $m_q \ll \Lambda_{\rm D}$ quark-like fermions. Unlike the Standard Model, the absence of a $SU(2)_L$ interaction allows the heavy quarks to be stable. The dark matter density is dominated by the heavy quarks, which today are hadronized. 

The dark radiation after the phase transition is composed of light pions with masses far below the confining scale. Due to the residual gauge force, the dark radiation is highly viscous, which results in novel modifications of the CMB power spectrum. In particular, we find that this model can relax the Hubble tension while maintaining the preferred values of $S_8$. Performing a combined fit to the {\em Planck} $TT$, $EE$, $TE$ and lensing data \cite{Planck:2018vyg, 2022JCAP...09..039C}, DESI DR2 \cite{2025PhRvD.112h3515A}, Pantheon+ \cite{2022ApJ...938..110B}, SH0ES \cite{2021ApJ...908L...6R,2022ApJ...934L...7R}, Big Bang Nucleosynthesis (BBN) data \cite{2018ApJ...855..102C, 2008CoPhC.178..956P, 2016PhRvL.116j2501M, 2009PhR...472....1I}, and ACT DR6 lensing \cite{2024ApJ...962..113M, 2024ApJ...962..112Q} we find our model is compatible with the combined cosmological data at the $\sim 3\sigma$ level, competitive with the ``Finalist'' models in the Hubble World Cup rankings \cite{Schoneberg:2026eys,Schoneberg:2026vaf}. In addition to accommodating the CMB data at this level, our model fits the DESI baryon acoustic oscillation (BAO) features better than $\Lambda$CDM using either the parameters derived from {\em Planck} measurements of the CMB alone or a joint fit to CMB and BAO data. The best-fit parameters for our model are in conflict with Lyman-$\alpha$ measurements \cite{McDonald:1999dt,eBOSS:2018qyj,Pedersen:2019ieb,Sabti:2021unj,Pedersen:2020kaw,Goldstein:2023gnw} of the small-scale matter power spectrum. However slightly larger values of $\Lambda_{\rm D}$ resolve this while still easing the Hubble Tension. 

Late-time self-interactions would seem to severely constrain this model, as the dark matter has a total (velocity-independent) self-interaction cross section of $\Lambda_{\rm D}^{-2}$, which can saturate the Bullet Cluster bounds \cite{Clowe:2006eq,Randall:2008ppe}. However, the cross section is misleading, as most dark matter-dark matter interactions in a present-day halo would result in the transfer of only ${\cal O}(\Lambda_{\rm D})$ energy. The majority of the mass is carried in the heavy quark, which must undergo a hard scattering to transfer significant momentum. Such hard scatters occur at perturbative scales with a much smaller cross section. Our model also contains new mechanisms for observationally-allowed dark matter cooling and energy transport which could modify dark matter structure at small scales.

We describe the model in Section~\ref{sec:model}. The cosmological effects of this form of dark matter and dark radiation on the CMB are described in Section~\ref{sec:cosmology}. Our analysis of cosmological data and the resulting model fit is described in Section~\ref{sec:data}. The scattering in late-Universe dark matter halos is described in Section~\ref{sec:presentday}, and we conclude in Section~\ref{sec:conclusions}.

\section{Low-Scale Dark QCD \label{sec:model}}

In our model, the dark sector is composed of fermions (akin to Standard Model quarks) charged under an $SU(N_c)$ gauge group that confines at a scale $\Lambda_{\rm D}$. The vector-like dark quarks are divided into $N_H$ heavy states $Q$ with masses $\gg \Lambda_{\rm D}$ and $N_\ell$ light states $q$ with masses $\ll \Lambda_{\rm D}$. Above the confinement scale, the free Lagrangian is
\begin{equation}
    {\cal L} \supseteq \sum_{j = 1}^{N_H} \bar{Q}_j \left( i  \slashed{D} - m_{H_j} \right) Q_j + \sum_{k = 1}^{N_\ell} \bar{q}_k  \left( i\slashed{D}  - m_{q_k} \right) q_k  - \frac{1}{4} G_{\mu\nu}^a G^{a\mu\nu}.
\end{equation}
At one loop, the gauge coupling running as a function of energy $\mu$ is
\begin{equation}
    \alpha_{\rm D}(\mu)^{-1} = \frac{1}{2\pi}\left( \frac{11}{3}N_c - \frac{2}{3}N_f \right) \ln\left(\frac{\mu}{\Lambda_{\rm D}}\right),
\end{equation}
where $N_f$ is the total number of dark quarks with mass less than $\mu$ and $\Lambda_{\rm D}$ is the confinement scale where $\alpha_{\rm D} \to \infty$.
For specificity, we will assume $N_c = 3$, one heavy quark $Q$ with mass $m_Q$, and two light quarks $u$ and $d$ with a common mass $m_q$. We assume no equivalents to the hypercharge and $SU(2)_L$ gauge groups, and the heavy quark is stable. Other choices are of course possible, and would modify some of the details of the fit to the cosmological data as discussed in Sections~\ref{sec:cosmology} and \ref{sec:data}.
The dark matter production mechanism is not the focus of this paper, though we note that asymmetric production \cite{Nussinov:1985xr,Petraki:2013wwa} is a viable scenario for the model.

At energies below $\Lambda_{\rm D}$, the vacuum state breaks the chiral symmetry of the light quarks, resulting in $N_\ell^2-1$ Nambu-Goldstone bosons which we refer to as ``dark pions'' in analogy with the Standard Model QCD.  The low-energy dynamics of these dark pions are described by Chiral Perturbation Theory ($\chi$PT). Working in our specific choice of $\ell = u,d$, we parameterize the Goldstone manifold using the unitary matrix field $\Sigma(x)$, which transforms as $\Sigma \to L \Sigma R^\dagger$ under $SU(2)_L \times SU(2)_R$:
\begin{equation}
\Sigma(x) = \exp \left( \frac{2i \Pi(x)}{f_{\rm D}} \right), \quad \Pi = \begin{pmatrix} \frac{\pi^0}{\sqrt{2}} & \pi^+ \\ \pi^- & -\frac{\pi^0}{\sqrt{2}} \end{pmatrix}.
\end{equation}
Here, $f_{\rm D}$ is the dark pion decay constant, which is set by the confinement scale, $f_{\rm D} \approx \Lambda_{\rm D} / 4\pi$. The leading-order effective Lagrangian is given by
\begin{equation}
\mathcal{L}_{\rm eff} = \frac{f_{\rm D}^2}{4} \text{Tr} \left( \partial_\mu \Sigma \partial^\mu \Sigma^\dagger \right) + \frac{f_{\rm D}^2 B_0}{2} \text{Tr} \left( M \Sigma^\dagger + \Sigma M^\dagger \right) .
\end{equation}
The first term describes the kinetic energy and self-interactions of the pions. The second term introduces the explicit symmetry breaking due to the small but non-zero light quark masses $M = \text{diag}(m_u, m_d)$. This term generates a mass for the dark pions
\begin{equation}
m_{\pi_{\rm D}}^2 \approx B_0 (m_u + m_d) \propto \Lambda_{\rm D} m_q .
\end{equation}
Since we have assumed $m_q \ll \Lambda_{\rm D}$, it follows that the dark pions are significantly lighter than the confinement scale ($m_{\pi_d} \ll \Lambda_{\rm D}$). This separation of scales is crucial for the cosmological evolution. It ensures that even after confinement, the dark sector retains relativistic degrees of freedom (the dark pions) for the temperature range below $\Lambda_{\rm D}$ and above $m_{\pi_{\rm D}}$, maintaining a form of dark radiation that eventually becomes non-relativistic matter in the late Universe (assumed to be well after the formation of the CMB). For the remainder of this work, we treat the dark pions as effectively massless.

Above the confinement scale, the dark matter is free heavy quarks $Q$, with a bath of dark radiation composed of $u/\bar{u}/d/\bar{d}$ quarks and dark gluons with a temperature $\hat{T} \equiv \xi T$, with $T$ the Standard Model bath temperature and we have defined the temperature ratio $\xi$. As the temperature decreases, the running dark gauge coupling $\alpha_{\rm D}$ diverges. After confinement, the dark radiation is composed only of the light mesons with mass $m_{\pi_{\rm D}}^2 \sim \Lambda_{\rm D} m_q \ll \Lambda_{\rm D}^2$. The dark matter at late times is the tower of hadronic states containing a $Q$ quark, all with mass $m_\chi \approx m_Q +{\cal O}(\Lambda_{\rm D})$. 

The lightest bound state for the $Q$ is a CP-odd doublet of scalar bound-state mesons $P_a$, composed of $Q\bar{u}$ and $Q\bar{d}$. The spin-0 $P_a$ and spin-1 mesons $P_{a\mu}$ can be combined into CP-even and CP-odd multiplets
\begin{equation}
\begin{aligned}
H_a & = \frac{(1+\slashed{v})}{2} \left[ -\gamma^5 P_a + P_{a\mu}^*\gamma^\mu\right] \\
\bar{H}_a & = \gamma^0 H_a \gamma^0 =  \left[ \gamma^5 P_a^* + \gamma^\mu P_{a\mu}\right] \frac{(1+\slashed{v})}{2}.
\end{aligned}
\end{equation}

At temperatures low enough to be in the confining phase, the velocities of the heavy quarks are small. We can therefore treat the four-velocity $v^\mu$ of heavy hadrons as a small parameter \cite{Casalbuoni:1996pg} for the purposes of perturbation theory. The leading terms in the heavy meson Lagrangian are
\begin{eqnarray*}
{\cal L} & \supseteq & i {\rm Tr} [ H_b v^\mu (\partial_\mu \delta_{ab} +V_{\mu ab}) \bar{H}_b] +ig {\rm Tr} [ H_b \gamma_\mu \gamma_5 A^\mu_{ab} \bar{H}_b],
\end{eqnarray*}
where
\begin{equation}
\begin{aligned}
V_\mu & = \frac{1}{2f_{\rm D}^2} \left[ \Pi,\partial_\mu \Pi\right] + {\cal O}(\Pi^4) \\
A_\mu & = -\frac{1}{f_{\rm D}} \partial_\mu \Pi + {\cal O}(\Pi^3).
\end{aligned}
\end{equation}

Having defined the interaction Lagrangians in both the free and confined phases, we now turn the behavior of the dark sector in the expanding Universe.

\section{Dark QCD Cosmology}
\label{sec:cosmology}

This model of dark matter coupled to a dark sector containing strongly self-interacting dark radiation modifies cosmological evolution around the release of the CMB photons at both the background level (though the addition of extra relativistic species), and at the perturbation level (by introducing perturbations in the dark radiation which couple to dark matter perturbations). We first consider the evolution of the bulk dark matter and dark radiation as a smooth background with a single temperature, and then turn to the perturbations.

\subsection{Thermal History \label{sec:thermal}}

The dark sector radiation bath at temperature $\hat{T}$ is characterized by its energy density, pressure, and entropy density:
\begin{equation}
    \rho_{\rm D} = \frac{\pi^2}{30} g_{\rho,{\rm D}}(\hat{T}) \hat{T}^4, \quad P_{\rm D} = \frac{\pi^2}{90} g_{P,{\rm D}}(\hat{T})  \hat{T}^4, \quad s_{\rm D} = \frac{2\pi^2}{45} g_{s,{\rm D}}(\hat{T})  \hat{T}^3. \label{eq:density_pressure_entropy}
\end{equation}
These are not independent, as $s_{\rm D}\hat{T} = (\rho_{\rm D}+P_{\rm D})$ and the effective number of degrees of freedom are related by $g_{s,{\rm D}} = \tfrac{1}{4}(3g_{\rho,{\rm D}}+g_{P,{\rm D}})$. In cosmology, the extra relativistic degrees of freedom are typically parametrized in terms of an excess in relativistic neutrino species $\Delta N_{\rm eff}$. For a dark sector temperature ratio $\xi = \hat{T}/T$ relative to the Standard Model bath,
\begin{equation}
\Delta N_{\rm eff}(T) = \frac{4}{7} \left( \frac{11}{4} \right)^{4/3} g_{s,{\rm D}}(T) \xi(T)^4.
\end{equation}

The radiation equation of state is defined as
\begin{equation}
    w \equiv \frac{P_{\rm D}}{\rho_{\rm D}} = \frac{1}{3} \frac{g_{P,{\rm D}}}{g_{\rho,{\rm D}}}.
\end{equation}
In the conformal limit, the three measured of degrees of freedom are identical and $w = 1/3$ -- as expected for radiation. The adiabatic sound speed is defined as $c_s^2 \equiv dP_{\rm D}/d\rho_{\rm D}$, and so is related to the evolution of the dark sector temperature with respect to the cosmic scale factor $a$:
\begin{equation}
    \frac{d\ln\hat{T}}{d\ln a} = - \left(1+\frac{\hat{T}}{3 g_{s,{\rm D}}(\hat{T})}\frac{dg_{s,\rm D}}{d\hat{T}}\right)^{-1} = -3c_s^2(\hat{T}).
\end{equation}
This recovers the standard expectation (temperature decreases as the inverse of $a$) when the number of relativistic degrees of freedom is constant, and results in the dark sector's temperature falling more slowly than $1/a$ when the number of relativistic species decreases.

Such a decrease occurs during the dark QCD phase transition at $\hat{T} \sim \Lambda_{\rm D}$. Above the transition (but below $m_Q$), there are 
\begin{equation}
    g_{s,{\rm D}}^{\rm UV} = \tfrac{7}{8} \times 4\times N_c\times N_\ell+2\times (N_c^2-1)
\end{equation}
effective relativistic degrees of freedom from the dark gluons and light quarks. In the IR ($\hat{T} <\Lambda_{\rm D})$, only the light pions are active and $g_{s,{\rm D}}$ falls to 
\begin{equation}
    g_{s,{\rm D}}^{\rm IR} = N_\ell^2-1.
\end{equation}
For the model parameters we assume in this paper ($N_c = 3$, $N_\ell = 2$), there are 37 degrees of freedom in the UV and 3 in the IR. Integrating across the entire phase transition, the dark sector temperature appears to ``reheat'' (relative to the temperature it would be expected to have due to the expansion of the Universe). Including possible reheating in the Standard Model sector, the temperature ratios $\xi = \hat{T}/T$ above and below the phase transition are related by
\begin{equation}
    \xi^{\rm IR} =  \left(\frac{g_{s,{\rm SM}}^{\rm IR}}{g_{s,{\rm SM}}^{\rm UV}} \frac{g_{s,{\rm D}}^{\rm UV}}{g_{s,{\rm D}}^{\rm IR}}\right)^{1/3}\xi^{\rm UV}.
\end{equation}
Any  change in Standard Model relativistic species occurring over the same temperature range is encapsulated in $g_{s,{\rm SM}}^{\rm IR}$ and $g_{s,{\rm SM}}^{\rm UV}$. With the benchmark values of $N_c$, $N_H$, and $N_\ell$ (and assuming no Standard Model degrees of freedom are becoming non-relativistic at $T \sim \Lambda_{\rm D}/\xi$), the dark sector reheats by $(37/3)^{1/3} \sim 2.31$ during the phase transition.

\begin{figure}[t!]
\centering
\includegraphics[width=0.7\textwidth]{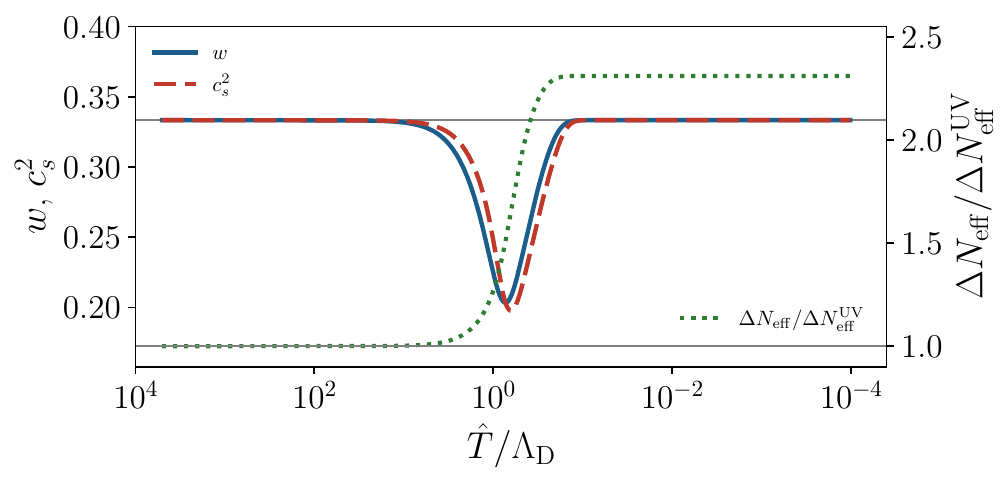}
\caption{The evolution of the dark radiation equation of state $w$ (solid blue), sound speed $c_s^2$ (dashed red), and $\Delta N_{\rm eff}$ (dotted green) through the phase transition at $\hat{T} = \Lambda_{\rm D}$. The vertical axis for $w$ and $c_s^2$ is on the left, and the axis for $\Delta N_{\rm eff}/\Delta N_{\rm eff}^{\rm UV}$ is on the right. We assume $N_c = 3$ and $N_\ell = 2$. \label{fig:step}}
\end{figure}

We parametrize the change in thermodynamic quantities during the phase transition using a Borsanyi prescription of the dimensionless trace anomaly
\begin{equation}
    \Delta(\hat{T}) \equiv \frac{\rho_{\rm D}- 3P_{\rm D}}{\hat{T}^4} = \frac{\pi^2}{90}\hat{T} \frac{dg_{P,{\rm D}}}{d\hat{T}},
\end{equation}
which has an analytic form fit to lattice QCD results \cite{Borsanyi:2013bia,HotQCD:2014kol}. 
The resulting non-conformal evolution of the equation of state and sound speed and reheating of the dark sector are shown in Figure~\ref{fig:step}.

This increase in the amount of radiation in the Universe increases the Hubble parameter $H(z)$ relative to the $\Lambda$CDM expectation. This decreases the sound horizon $r_s$ at the redshift of last scattering $z_*$
\begin{equation}
    r_s = \int_{z_*}^\infty \frac{c_s(z)}{H(z)}dz.
\end{equation}
(Note that $c_s$ here is the sound speed of the photon-electron plasma, not of the dark sector.)
As {\em Planck} measures the angular scale of the acoustic oscillations $\theta_* = r_s/D_M(z_*)$ very precisely, reducing $r_s$ requires decreasing the comoving distances to the surface of last scattering. This can be accomplished by increasing the Hubble parameter in the late Universe \cite{Knox:2019rjx}, and so provides a mechanism to ease the Hubble tension. We discuss the detailed fits to data in Section~\ref{sec:data}, where we find that the values of $\Lambda_{\rm D}$ that provides the most improvement to the Hubble tension are $\sim {\rm MeV}$. As a result, the step in dark radiation occurs at horizons much smaller than the critical scales for the CMB. Our resolution to the tension is not therefore {\em primarily} driven by the step in the number of effective relativistic degrees of freedom, but rather by the behavior of the dark radiation itself as we discuss next.

\subsection{Perturbations}

In addition to the evolution of the dark radiation background, cosmological data is sensitive to the dark matter-dark radiation perturbations, through the gravitational coupling of the photons to other forms of energy. To calculate the observational constraints on the dark QCD model therefore requires a treatment of the evolution of these perturbations. In particular, this model couples the dark matter to a highly viscous radiation fluid, causing the perturbations to evolve in unique ways compared to free-streaming neutrinos or the perfect fluid of tightly-coupled photons.

To understand the cosmological consequences of the dark matter-dark radiation interactions, it is useful to consider a heavy dark matter particle propagating through the thermal bath at temperatures much below its mass $m_\chi \gg \Lambda_{\rm D} \sim \hat{T}$. Each dark matter particle must absorb many light kicks from the dark radiation before its momentum changes appreciably. It therefore undergoes Brownian motion. This is the case in all epochs of relevance in calculating the CMB power spectrum, and applies both above and below the confinement scale. The collision equation reduces to a Fokker-Planck operation with a drag coefficient $\Gamma_{\rm drag}$ and momentum-space diffusion coefficient $\kappa$ \cite{Svetitsky:1987gq,Moore:2004tg}, which are related by
\begin{equation}
    \Gamma_{\rm drag} = \frac{\kappa}{2m_\chi \hat{T}}.
\end{equation}
The function $\kappa(\hat{T})$ (or alternatively, $\Gamma_{\rm drag}$) sets the behavior of energy and momentum exchange, while the shear is set by the shear viscosity function $\eta(\hat{T})$. 

The drag rate can be calculated from the theory of Section~\ref{sec:model} in both the free and confined phases. In the free theory,
\begin{equation}
    \Gamma_{\rm drag}^{\rm UV} = \frac{\zeta(3) (N_c^2-1)}{\pi} \frac{\alpha_{\rm D}^2 \hat{T}^2}{m_\chi} \ln \alpha_{\rm D}^{-1}.
\end{equation}
As the Hubble parameter $H(T) \propto T^2$ in the radiation-dominated Universe, the ratio $\Gamma_{\rm drag}^{\rm UV}/H$ is constant (up to the running of $\alpha_{\rm D}$). Thus, assuming the dark matter is strongly coupled to the dark radiation, it will never decouple in the free phase.
After confinement, the dark pion-dark matter drag rate can be calculated using chiral and heavy quark effective perturbation theories \cite{Wise:1992hn,Casalbuoni:1996pg}, becoming
\begin{equation}
    \Gamma_{\rm drag}^{\rm IR} \approx \frac{\hat{T}^6}{m_\chi f_{\rm D}^4}.
\end{equation}
Thus, after confinement the drag rate collapses and eventually the dark matter will decoupled from the dark radiation. However, this will occur more gradually than in the scenarios considered in \cite{Aloni:2021eaq,Joseph:2022jsf}. 

The dark radiation (in both the confined and free phases) will have self-interactions. The self-interaction cross section for dark pions is
\begin{equation}
    \sigma_{\pi\pi} \approx \frac{\hat{T}^2}{f_{\rm D}^4}.    
\end{equation} 
Critically for the model's effect on the CMB, in the confined phase, the dark pions are not just self-interacting, they act as a viscous imperfect fluid. The viscosity is approximately the radiation density $\rho_\pi$ times the mean free path length $1/n_\pi\sigma_{\pi\pi}$:
\begin{equation}
    \eta_{\rm visc} \approx \frac{f_{\rm D}^4}{\hat{T}}.
\end{equation}
This defines a relaxation time for the dark radiation fluid
\begin{equation}
    \tau_{\rm relax} = \frac{5\eta_{\rm visc}}{(1+w)\rho_{\rm DR}}. \label{eq:relax_tau}
\end{equation}
Note that the dark radiation becomes more free-streaming as the Universe cools.

The heavy dark matter kinetically decouples when the drag rate falls to the Hubble rate \cite{Bertschinger:2006nq,Bringmann:2006mu}:
\begin{equation}
\Gamma_{\rm drag}(\hat{T}^{\rm kd})=H(T^{\rm kd}).
\label{eq:kd-condition-mt}
\end{equation}
Inserting the steeply-falling drag rate after confinement with $\hat{T}=\xi_{\rm kd} T$, we find kinetic decoupling occurs at
\begin{equation}
T^{\rm kd}=\left(\frac{21\times 1.66\sqrt{g_\ast}}{\pi^3}\,\frac{m_\chi f_{\rm D}^4}{\xi_{\rm kd}^6 M_{\rm Pl}}\right)^{1/4}.
\label{eq:kd-mt}
\end{equation}
The decoupling redshift is therefore only weakly dependent on the dark matter mass: 
\begin{equation}
z_{\rm kd}\simeq1.1\times10^{5}\left(\frac{\Lambda_{\rm D}}{{\rm MeV}}\right)\left(\frac{m_\chi}{{\rm TeV}}\right)^{1/4}\left(\frac{\xi_{\rm kd}}{0.52}\right)^{-3/2}  \;\propto\;\Lambda_{\rm D}\,m_\chi^{1/4}\,\left(\Delta N_{\rm eff}^{\rm IR}\right)^{-3/8}.
\label{eq:zkdscaling}
\end{equation}
Eq.~\eqref{eq:zkdscaling} is the \emph{only} way $m_\chi$ enters the linear cosmology, while $\Lambda_{\rm D}$ enters only here and in the relaxation time Eq.~\eqref{eq:relax_tau} (as $f_{\rm D}\propto \Lamd$).
As a result, the three-dimensional parameter space $(\Delta_{\rm eff}^{\rm IR},\Lambda_{\rm D},m_\chi)$ is largely degenerate along the direction
that holds $\zkd$ fixed.

We implement the changes in the evolution of the early Universe as a result of the dark QCD model (DQCD) by modifying the \CLASS\, \cite{Diego_Blas_2011} package, working in the synchronous gauge. The background temperature of the dark sector evolves as described in Section~\ref{sec:thermal}, and the average densities of dark radiation and dark matter modify the Hubble parameter at the level of the Friedmann Equations. The conformal Hubble parameter is $\mathcal{H}$ and primes denote derivatives with respect to conformal time. In the following $h$ is the trace of the metric perturbation in the synchronous gauge, not to be confused with the dimensionless Hubble constant. The trace-free metric perturbation is $\eta$.

For numerical stability, we treat the regime where dark matter and dark radiation are tightly-coupled separately rather than smoothly transitioning. There are two tight-coupling condition, one for the drag and one for viscosity. The drag tight-coupling condition is $(1+R_\chi)\Gamma_{\rm drag} \gg H$ and $\gg k/a$ (here $R_\chi = \rho_\chi/(1+w)\rho_{\rm DR}$). The viscosity tight coupling condition is $\tau_{\rm relax}^{-1} \gg H,\, k/a$.

When the two fluids are tightly coupled, the dark matter and dark radiation density contrasts, velocity divergences, dark radiation anisotropic stress, and higher moments for the dark radiation evolve as
\begin{equation}
\begin{aligned}
    \delta_\chi' & =  -\theta_\chi - \tfrac{1}{2}h'\\
    \delta_{\rm DR}' & = -(1+w)(\theta_{\rm DR}+\tfrac{1}{2}h') - 3\mathcal{H}(c_s^2 - w)\,\delta_{\rm DR} \\
    \theta_\chi'&  = \frac{1}{1+R_\chi}\left(-R_\chi{\mathcal H} \theta_\chi -(1-3c_s^2)\mathcal{H}\theta_{\rm DR} +\frac{c_s^2 k^2}{1+w}\delta_{\rm DR} + S_v'\right) \\
    \theta_{\rm DR}' & =  \theta_\chi' - S_v' \\
    \sigma_{\rm DR} & =  \frac{4}{15}\frac{\tau_{\rm relax}}{a}\Big(\theta_{\rm DR} + \tfrac12h' + 3\eta'\Big) \\
    F_{{\rm DR},\ell\ge3} & = 0.
\end{aligned}
\end{equation}
Here, the slip is $S_v\equiv\theta_\chi-\theta_{\rm DR}$  and 
\begin{equation}
S_v' = \frac{1}{(1+R)a\Gamma_{\rm drag}}\left[-(\mathcal H' + \mathcal H^2)\theta_\chi- \tfrac12\mathcal H k^2\delta_{\rm DR}+ c_s^2 k^2(\theta_{\rm DR} + \tfrac{1}{2}h') \right].
\end{equation}

In the weakly coupled regimes, the density contrast equations remain unchanged, while
\begin{equation}
\begin{aligned}
    \theta_\chi' & = -{\mathcal H}\theta_\chi + a \Gamma_{\rm drag}(\theta_{\rm DR} - \theta_\chi) \\
    \theta_{\rm DR}' & = -(1-3c_s^2)\mathcal H\theta_{\rm DR} + k^2\Big(\frac{c_s^2}{1+w}\delta_{\rm DR} - \sigma_{\rm DR}\Big) + aR\,\Gamma_{\rm drag}(\theta_\chi-\theta_{\rm DR}) \\
    \sigma_{\rm DR}'&  = \frac{4}{15}\Big(\theta_{\rm DR}+\tfrac12h'+3\eta'\Big) - \frac{3}{10}kF_{{\rm DR},3} - \frac{a}{\tau_{\rm relax}}\sigma_{\rm DR} \\
    F_{{\rm DR},\ell}' & = \frac{k}{2\ell+1}\big[\ell F_{{\rm DR},\ell-1} - (\ell+1)F_{{\rm DR},\ell+1}\big] - \frac{a}{\tau_{\rm relax}}F_{{\rm DR},\ell}.
\end{aligned}
\end{equation}
The drag tight coupling condition controls the switch between the two sets of equations for the density contrasts and velocities, while the anisotropic stress and higher moments depend on the viscosity tight-coupling condition.

\begin{figure}[t!]
\centering
\includegraphics[width=\textwidth]{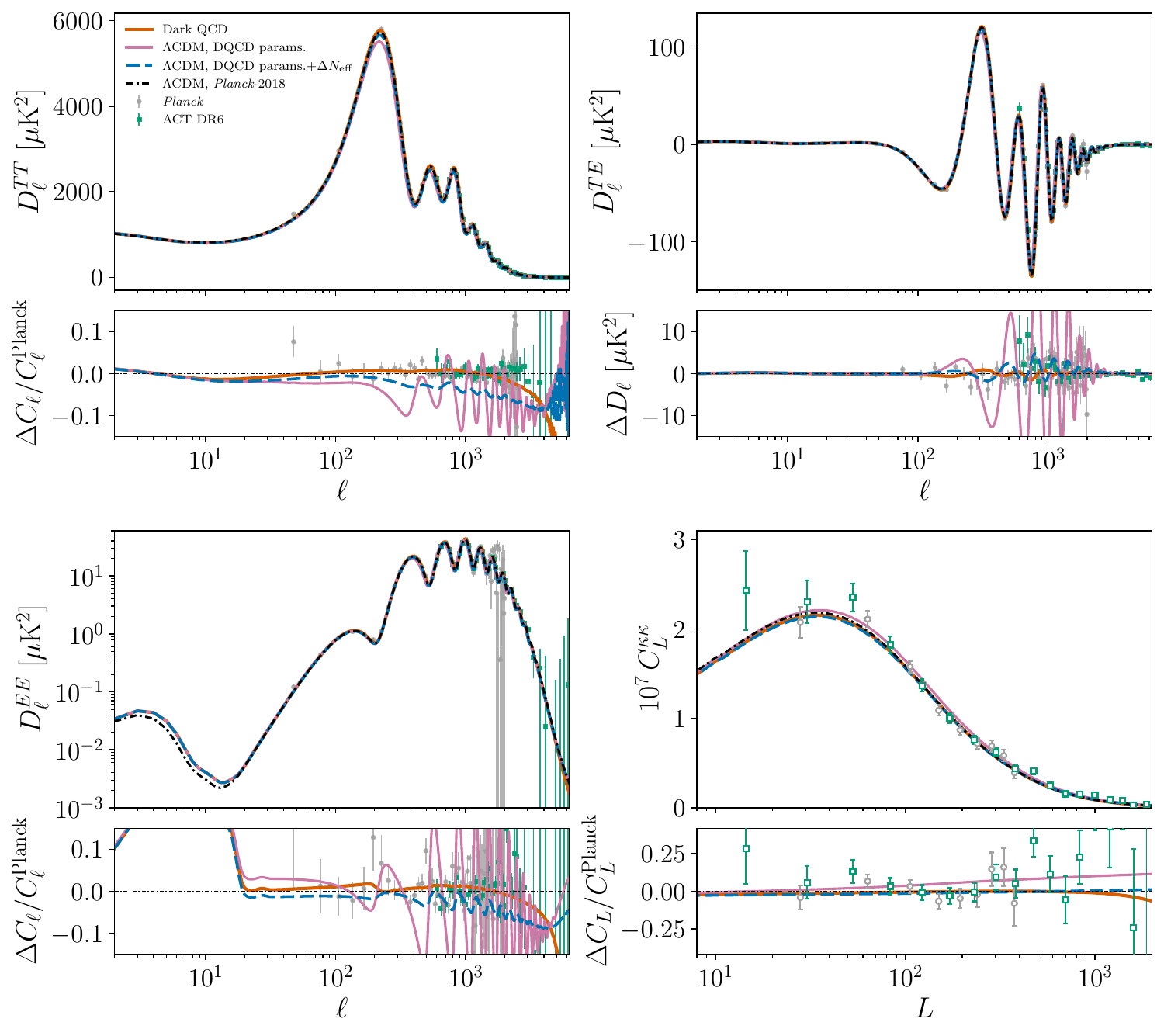}
\caption{Example CMB $TT$ (upper left), $TE$ (upper right), $EE$ (lower right), and lensing (lower right) power spectra for the dark QCD model using the maximum likelihood cosmological parameters of Table~\ref{tab:fits} when fit to all data sets (red solid), $\Lambda$CDM using these same cosmological parameters (pink solid), $\Lambda$CDM using these parameters with an additional form of free-streaming $\Delta N_{\rm eff} = 0.577$ (blue dashed), and $\Lambda$CDM using the best-fit cosmology found by {\em Planck} \cite{Planck:2018vyg} (black dashed-dotted). The deviations in the lower panels are taken with respect to the best-fit {\em Planck} cosmology. The {\em Planck} 2018 data \cite{Planck:2018vyg} is shown in gray, the ACT DR6 data \cite{2024ApJ...962..113M, 2024ApJ...962..112Q} in teal. The data and the fitting procedure are discussed in Section~\ref{sec:data}. \label{fig:example_CMB}}
\end{figure}

In Figure~\ref{fig:example_CMB} we show the CMB multipoles as calculated by \CLASS\, using our DQCD model with parameters set to maximize the likelihood of the cosmological datasets as described in Section~\ref{sec:data}. In particular, $m_\chi = 959$~GeV, $\Lambda_{\rm D} = 1.1$~MeV, $\Delta N_{\rm eff}^{\rm IR} = 0.577$, and $H_0 = 71.82~{\rm km/s/Mpc}$.
We compare against the best-fit {\em Planck} $\Lambda$CDM parameters \cite{Planck:2018vyg}, $\Lambda$CDM assuming the same cosmological parameters as in our  DQCD model, and $\Lambda$CDM with those parameters and an additional form of free-streaming dark radiation with the same $\Delta N_{\rm eff}^{\rm IR}$ as in DQCD. As can be seen, the viscous nature of the dark radiation suppresses oscillations at high $\ell$ relative to free-streaming relativistic particles. Additionally, while increasing $H_0$ in $\Lambda$CDM with or without extra free-streaming relativistic particles results in significant deviations in the power spectrum at high $\ell$, they are again suppressed in DQCD.

\section{Comparison with Data \label{sec:data}}

Our dark QCD model has three parameters that affect cosmological observables. We take these to be the mass of the dark matter $m_\chi$, the confinement scale $\Lambda_{\rm D}$, and the number of additional relativistic degrees of freedom after confinement $\Delta N_{\rm eff}^{\rm IR}$. As the cosmological observables depends only weakly with the dark matter mass, we adopt a canonical choice of $m_\chi = 959$~GeV. We then calculate the likelihood of the cosmological data in the dark QCD model when varying $\Lambda_{\rm D}$ and $\Delta N_{\rm eff}^{\rm IR}$ along with six parameters of the standard $\Lambda$CDM: $A_{\rm s}$, $n_{\rm s}$, $\theta_{\rm s}$, $\omega_{\rm b}$, $\omega_{\rm CDM}$, $\tau_{\rm reio}$. When combined with the other parameters, the angular size of the sound horizon $\theta_{\rm s}$ can be converted to $H_0$, and we express our results in terms of the latter. We compare the likelihood of our new physics model to the likelihood of the data under $\Lambda$CDM.

In the likelihood analyses, we consider the following datasets:
\begin{itemize}
    \item The 2018 {\em Planck} analysis of CMB $TT$, $EE$, $TE$, and lensing data \cite{Planck:2018vyg, 2022JCAP...09..039C}.
    \item CMB lensing measurements from ACT DR6 \cite{2024ApJ...962..113M, 2024ApJ...962..112Q}.
    \item The DESI DR2 \cite{2025PhRvD.112h3515A} measurements of BAO angular scales.
    \item The Pantheon+ \cite{2022ApJ...938..110B} supernovae data.
    \item The 2022 SH0ES 
    \cite{2022ApJ...934L...7R} measurements of supernovae.
    \item The BBN D/H abundance from Cooke {\em et al} \cite{2018ApJ...855..102C}. 
\end{itemize}
Except for the BBN constraints, we use the likelihood packages provided in the \textsc{Cobaya} cosmological Markov-Chain Monte Carlo (MCMC) sampler code \cite{Torrado:2020dgo}. For the deuterium BBN likelihood, we adopt a Gaussian likelihood around the Cooke \cite{2018ApJ...855..102C} measurement, evaluated over a lookup table pre-computed using \textsc{PArthENoPE} \cite{2008CoPhC.178..956P, 2016PhRvL.116j2501M, 2009PhR...472....1I, 2020Natur.587..210M}. This allows us to consistently apply BBN constraints even as the number of effective relativistic degrees of freedom evolves with temperature during to the dark phase transition. 

We explore the viable parameter space of dark QCD
using \textsc{Cobaya} 
wrapped around \textsc{Class} v3.3.4\footnote{\href{https://github.com/lesgourg/class_public}{https://github.com/lesgourg/class\_public}} 
\cite{Diego_Blas_2011} with non-linearities disabled for the $\Lambda$CDM likelihoods. We modify this version of \textsc{Class} as per the discussion in Section~\ref{sec:cosmology} to implement the dark QCD model.
The priors adopted for the eight sampled parameters ($\Lambda_{\rm D}$, $\Delta N_{\rm eff}^{\rm IR}$ and the six $\Lambda$CDM parameters) are documented in Table~\ref{tab:priors}. The eight MCMC chains were sampled to Gelman-Rubin convergence criteria of $R-1<0.2$, with a $30\%$ burn-in fraction removed. 

\begin{figure}
    \centering
    \includegraphics[width=0.9\textwidth]{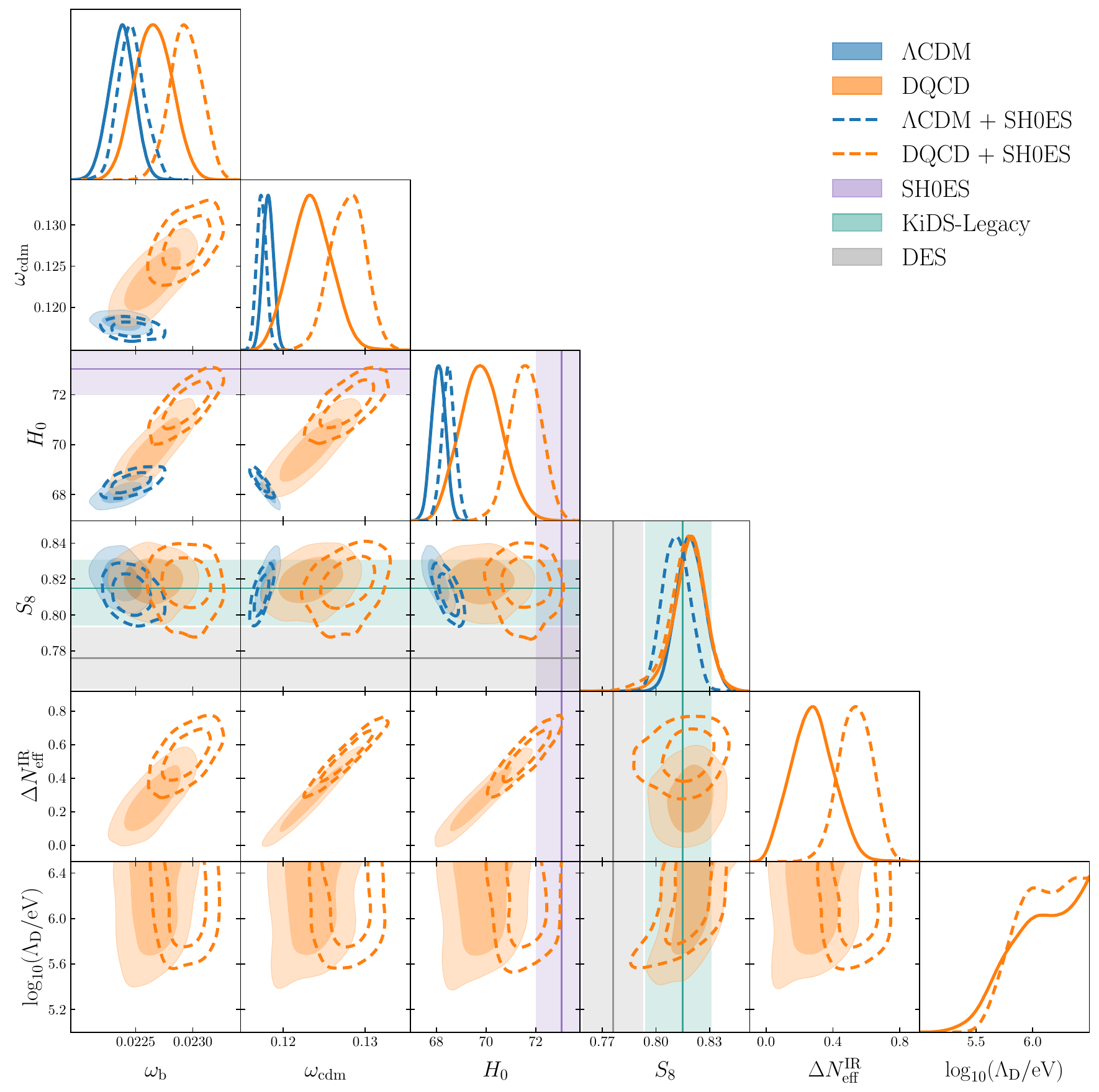}
    \caption{MCMC posteriors of {\em Planck}+DESI+SN+BBN+ACT data for $\Lambda$CDM (blue) and the dark QCD model (orange). The solid contours indicate the regions preferred without including SH0ES data, the dashed lines indicate the preference including these late-time supernovae measurements. The SH0ES measurement of $H_0$ \cite{2022ApJ...934L...7R} is shown in purple. We show measurements of $S_8$ from KiDS-Legacy \cite{Wright:2025xka} (teal) and DES \cite{DES:2026fyc} (gray). }
    \label{fig:likelihood_corner}
\end{figure}

In Table~\ref{tab:fits} we present the profile-likelihood best-fit and marginalized posterior means and 68\% confidence intervals for $\Lambda$CDM and dark QCD across multiple combinations of datasets. These posteriors are also shown in Figure~\ref{fig:likelihood_corner}. We include the derived marginalized posterior range and best-fit value for the derived cosmological parameter $S_8$ (the matter density clustering at $8~{\rm Mpc}/h$). In addition to the datasets included in our fits, $S_8$ has been measured by KiDS-Legacy \cite{Wright:2025xka} and DES  \cite{DES:2026fyc}.

\begin{table}[t]
    \centering
    \begin{tabular}{|l|c||l|c||l|c|} \hline
    \multicolumn{2}{|c||}{\textbf{$\Lambda$CDM}} & \multicolumn{2}{c||}{\textbf{Dark QCD}} & \multicolumn{2}{c|}{\textbf{Nuisance}} \\ \hline \hline
        $\ln(10^{10} A_s)$        & $[1.61,\, 3.91]$   & $\Delta N_{\rm eff}^{\rm IR}$         & $[10^{-5},\, 2.70]$ & $M_b$ & $[-20,\, -18]$ \\
        $n_s$                     & $[0.8,\, 1.2]$      & $\log_{10}(\Lambda_{\rm D}/{\rm eV})$ & $[5.0,\, 6.5]$      & & \\
        $100\,\theta_{\rm s}$     & $[0.5,\, 10]$       & & & & \\
        $\omega_{\rm b}$          & $[0.005,\, 0.1]$    & & & & \\
        $\omega_{\rm CDM}$        & $[0.001,\, 0.99]$   & & & & \\
        $\tau_{\rm reio}$         & $[0.01,\, 0.8]$     & & & & \\
        \hline
    \end{tabular}
        \caption{Priors on sampled parameters for the MCMC likelihood evaluations. All the $\Lambda$CDM parameters are varied in every fit.}
    \label{tab:priors}
\end{table}

As can be seen in Figure~\ref{fig:likelihood_corner} and in Table~\ref{tab:fits}, the dark QCD model posterior over the datasets including SH0ES finds $H_0 = 71.60\pm 0.63~{\rm km/s/Mpc}$, compared to the SH0ES-only measurement of $73.04\pm 1.04~{\rm km/s/Mpc}$. The posterior range for $\Lambda_{\rm D}$ is $1.3^{+1.0}_{-0.3}$~MeV. However the fit is relatively flat in this parameter, and there is a ``shelf'' of larger confinement scales which also have large values of $H_0$ and are compatible with the data. 

The dark QCD parameters preferred by the posterior are also consistent with the KiDS-Legacy measurement of $S_8$. These DQCD parameters are in tension with the DES measurement of $S_8$, but at the same level as found in $\Lambda$CDM. Note that neither KiDS-Legacy or DES are included when fitting the model parameters.

\begin{figure}
    \centering
    \includegraphics[width=0.5\textwidth]{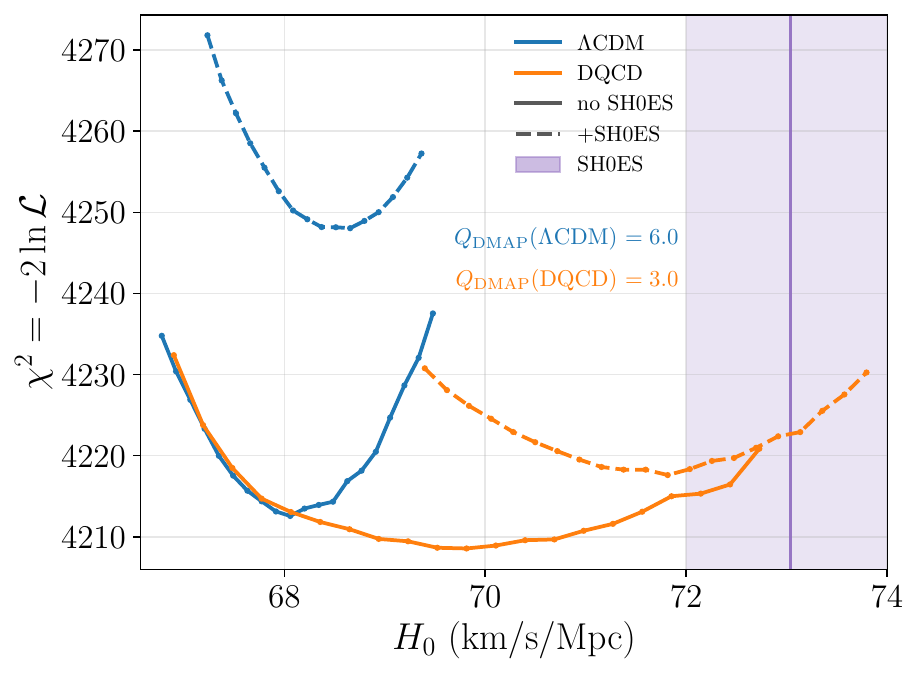}
    \caption{Frequentist likelihood profiling over $H_0$, compared between $\Lambda$CDM (orange) and dark QCD (blue) cosmology evaluated with the {\em Planck}+DESI+SN+BBN+ACT both without SH0ES data (solid lines) and with SH0ES (dashed lines), enabling a direct readout of the $Q_{\rm DMAP}$ value when the late-time supernovae measurements are included. The $1\sigma$ range of $H_0$ preferred by SH0ES is shown in purple.}
    \label{fig:hubble}
\end{figure}

In addition to the Bayesian posterior, we perform a frequentist likelihood profiling using the \textsc{Prospect} code \cite{Holm:2023uwa} which uses simulated annealing to optimize for the global best fit. In Figure~\ref{fig:hubble}, we show the $\chi^2$ of a profile over a grid of fixed $H_0$, warm-started from covariance matrices from the MCMC runs, and all other parameters optimized. Results are shown for $\Lambda$CDM and DQCD, both with and without the SH0ES measurement included. As can be seen, $\Lambda$CDM cannot easily accommodate the SH0ES measurement of $H_0$, while the dark QCD model can approach the SH0ES values with only a moderate degradation of the fit.

This frequentist analysis allows us to calculate the difference of the maximum a posteriori ($Q_{\rm DMAP}$) \cite{Raveri:2021wfz} for a combination of datasets $A$:
\begin{equation}
    Q_{\rm DMAP}^2 = \chi^2_{\rm min}(A+{\rm SH0ES}) - \chi_{\rm min}^2(A)- \chi_{\rm min}^2({\rm SH0ES}).
\end{equation}
This is a measure of the statistical improvement of a model when fit to datasets $A$ with and without the SH0ES measurement, and has been adopted by the Hubble World Cup analysis \cite{Schoneberg:2026eys,Schoneberg:2026vaf} as a figure of merit to evaluate the success of a new physics model in reducing the Hubble tension. Note that the $\chi_{\rm min}^2({\rm SH0ES})$ term is zero, since this represents a one-parameter fit that both $\Lambda$CDM and dark QCD can minimize exactly. $\Lambda$CDM has a $Q_{\rm DMAP} = 6.0\sigma$ when comparing across the {\em Planck}, ACT, DESI, Pantheon+, and BBN datasets, while the dark QCD model has $Q_{\rm DMAP}=3.0\sigma$. By this metric, the dark QCD model appears competitive with the ``finalist'' models in the Hubble World Cup, though of course this is not a substitute for the full statistical analysis using identical code.

\begin{table*}
    \centering
    \resizebox{\textwidth}{!}{%
    \begin{tabular}{l l c c c c c c c c c c c }
        \hline\hline
        Data & Model & $\omega_\mathrm{b}$ & $\omega_\mathrm{cdm}$ & $\log(10^{10}A_\mathrm{s})$ & $n_\mathrm{s}$ & $\tau_\mathrm{reio}$ & $H_0$ & $S_8$ & $\Delta N_{\rm eff}^\mathrm{IR}$ & $\log_{10}(\Lambda_{\rm D}/\mathrm{eV})$ & $\chi^2$ & $\Delta \chi^2$\\
        \hline
        \multirow{2}{*}{ \makecell{{\em Planck}+DESI+SN \\ (Baseline)} } & $\Lambda$CDM & \makecell{$0.02248^{+0.00012}_{-0.00012}$ \\ $(0.02256)$} & \makecell{$0.1179^{+0.0006}_{-0.0006}$ \\ $(0.1176)$} & \makecell{$3.054^{+0.014}_{-0.015}$ \\ $(3.056)$} & \makecell{$0.9693^{+0.0034}_{-0.0034}$ \\ $(0.9705)$} & \makecell{$0.0611^{+0.0068}_{-0.0078}$ \\ $(0.0606)$} & \makecell{$68.27^{+0.27}_{-0.29}$ \\ $(68.46)$} & \makecell{$0.812^{+0.008}_{-0.008}$ \\ $(0.809)$} & --- & --- & 4194.2 & 0 \\
         & DQCD & \makecell{$0.02266^{+0.00015}_{-0.00017}$ \\ $(0.02257)$} & \makecell{$0.1219^{+0.0020}_{-0.0030}$ \\ $(0.1201)$} & \makecell{$3.050^{+0.015}_{-0.015}$ \\ $(3.051)$} & \makecell{$0.9700^{+0.0034}_{-0.0035}$ \\ $(0.9697)$} & \makecell{$0.0609^{+0.0070}_{-0.0074}$ \\ $(0.0602)$} & \makecell{$69.54^{+0.66}_{-0.99}$ \\ $(68.96)$} & \makecell{$0.809^{+0.014}_{-0.008}$ \\ $(0.811)$} & \makecell{$0.210^{+0.078}_{-0.185}$ \\ $(0.117)$} & \makecell{$5.95^{+0.55}_{-0.17}$ \\ $(5.78)$} & 4193.6 & 0 \\
        \hline
        \multirow{2}{*}{Baseline+SH0ES} & $\Lambda$CDM & \makecell{$0.02262^{+0.00011}_{-0.00013}$ \\ $(0.02257)$} & \makecell{$0.1170^{+0.0006}_{-0.0006}$ \\ $(0.1174)$} & \makecell{$3.059^{+0.014}_{-0.016}$ \\ $(3.057)$} & \makecell{$0.9717^{+0.0033}_{-0.0034}$ \\ $(0.9722)$} & \makecell{$0.0639^{+0.0069}_{-0.0080}$ \\ $(0.0626)$} & \makecell{$68.75^{+0.27}_{-0.28}$ \\ $(68.55)$} & \makecell{$0.804^{+0.007}_{-0.008}$ \\ $(0.808)$} & --- & --- & 4226.7 & 32.5 \\
         & DQCD & \makecell{$0.02294^{+0.00015}_{-0.00014}$ \\ $(0.02295)$} & \makecell{$0.1280^{+0.0024}_{-0.0024}$ \\ $(0.1278)$} & \makecell{$3.045^{+0.014}_{-0.015}$ \\ $(3.046)$} & \makecell{$0.9720^{+0.0035}_{-0.0034}$ \\ $(0.9723)$} & \makecell{$0.0608^{+0.0070}_{-0.0071}$ \\ $(0.0602)$} & \makecell{$71.69^{+0.68}_{-0.68}$ \\ $(71.69)$} & \makecell{$0.804^{+0.021}_{-0.008}$ \\ $(0.815)$} & \makecell{$0.547^{+0.115}_{-0.116}$ \\ $(0.543)$} & \makecell{$5.98^{+0.52}_{-0.18}$ \\ $(6.26)$} & 4203.4 & 9.8 \\
        \hline
        \multirow{2}{*}{Baseline+BBN} & $\Lambda$CDM & \makecell{$0.02237^{+0.00011}_{-0.00012}$ \\ $(0.02234)$} & \makecell{$0.1180^{+0.0006}_{-0.0007}$ \\ $(0.1183)$} & \makecell{$3.052^{+0.014}_{-0.014}$ \\ $(3.051)$} & \makecell{$0.9685^{+0.0031}_{-0.0035}$ \\ $(0.9685)$} & \makecell{$0.0600^{+0.0070}_{-0.0074}$ \\ $(0.0585)$} & \makecell{$68.10^{+0.29}_{-0.27}$ \\ $(67.97)$} & \makecell{$0.814^{+0.008}_{-0.008}$ \\ $(0.817)$} & --- & --- & 4198.7 & 4.5  \\
         & DQCD & \makecell{$0.02263^{+0.00017}_{-0.00017}$ \\ $(0.02265)$} & \makecell{$0.1231^{+0.0025}_{-0.0026}$ \\ $(0.1229)$} & \makecell{$3.049^{+0.014}_{-0.016}$ \\ $(3.046)$} & \makecell{$0.9699^{+0.0036}_{-0.0037}$ \\ $(0.9711)$} & \makecell{$0.0607^{+0.0072}_{-0.0083}$ \\ $(0.0593)$} & \makecell{$69.76^{+0.80}_{-0.93}$ \\ $(69.77)$} & \makecell{$0.812^{+0.013}_{-0.009}$ \\ $(0.816)$} & \makecell{$0.267^{+0.129}_{-0.136}$ \\ $(0.261)$} & \makecell{$5.98^{+0.46}_{-0.21}$ \\ $(6.23)$} & 4194.8 & 1.2 \\
        \hline
        \multirow{2}{*}{ \makecell{Baseline+\\ BBN +SH0ES}} & $\Lambda$CDM & \makecell{$0.02247^{+0.00011}_{-0.00011}$ \\ $(0.02252)$} & \makecell{$0.1172^{+0.0006}_{-0.0007}$ \\ $(0.1169)$} & \makecell{$3.056^{+0.014}_{-0.015}$ \\ $(3.059)$} & \makecell{$0.9711^{+0.0034}_{-0.0035}$ \\ $(0.9723)$} & \makecell{$0.0630^{+0.0067}_{-0.0080}$ \\ $(0.0643)$} & \makecell{$68.54^{+0.28}_{-0.28}$ \\ $(68.67)$} & \makecell{$0.806^{+0.008}_{-0.008}$ \\ $(0.804)$} & --- & --- & 4233.4 & 39.2 \\
         & DQCD & \makecell{$0.02294^{+0.00014}_{-0.00013}$ \\ $(0.02298)$} & \makecell{$0.1278^{+0.0020}_{-0.0021}$ \\ $(0.1286)$} & \makecell{$3.045^{+0.014}_{-0.016}$ \\ $(3.042)$} & \makecell{$0.9722^{+0.0034}_{-0.0034}$ \\ $(0.9728)$} & \makecell{$0.0613^{+0.0065}_{-0.0082}$ \\ $(0.0597)$} & \makecell{$71.68^{+0.62}_{-0.61}$ \\ $(71.90)$} & \makecell{$0.806^{+0.018}_{-0.008}$ \\ $(0.815)$} & \makecell{$0.543^{+0.097}_{-0.096}$ \\ $(0.581)$} & \makecell{$6.03^{+0.47}_{-0.18}$ \\ $(6.28)$} & 4203.6 & 10.0\\
        \hline
        \multirow{2}{*}{ \makecell{Baseline+\\ ACT+BBN}} & $\Lambda$CDM & \makecell{$0.02238^{+0.00011}_{-0.00011}$ \\ $(0.02238)$} & \makecell{$0.1181^{+0.0006}_{-0.0007}$ \\ $(0.1182)$} & \makecell{$3.062^{+0.013}_{-0.015}$ \\ $(3.059)$} & \makecell{$0.9695^{+0.0034}_{-0.0033}$ \\ $(0.9702)$} & \makecell{$0.0638^{+0.0071}_{-0.0084}$ \\ $(0.0603)$} & \makecell{$68.08^{+0.28}_{-0.29}$ \\ $(68.06)$} & \makecell{$0.819^{+0.008}_{-0.008}$ \\ $(0.819)$} & --- & --- & 4212.6 & 18.4 \\
         & DQCD & \makecell{$0.02265^{+0.00016}_{-0.00017}$ \\ $(0.02265)$} & \makecell{$0.1234^{+0.0026}_{-0.0026}$ \\ $(0.1236)$} & \makecell{$3.057^{+0.013}_{-0.013}$ \\ $(3.055)$} & \makecell{$0.9710^{+0.0034}_{-0.0034}$ \\ $(0.9708)$} & \makecell{$0.0629^{+0.0071}_{-0.0076}$ \\ $(0.0608)$} & \makecell{$69.81^{+0.83}_{-0.84}$ \\ $(69.81)$} & \makecell{$0.819^{+0.009}_{-0.008}$ \\ $(0.822)$} & \makecell{$0.278^{+0.123}_{-0.135}$ \\ $(0.283)$} & \makecell{$6.07^{+0.43}_{-0.13}$ \\ $(6.36)$} & 4208.6 & 15.0\\
        \hline
        \multirow{2}{*}{ \makecell{Baseline+\\ ACT+BBN+SH0ES}} & $\Lambda$CDM & \makecell{$0.02246^{+0.00010}_{-0.00013}$ \\ $(0.02253)$} & \makecell{$0.1173^{+0.0006}_{-0.0006}$ \\ $(0.1171)$} & \makecell{$3.065^{+0.012}_{-0.015}$ \\ $(3.065)$} & \makecell{$0.9715^{+0.0032}_{-0.0030}$ \\ $(0.9726)$} & \makecell{$0.0654^{+0.0068}_{-0.0082}$ \\ $(0.0645)$} & \makecell{$68.49^{+0.27}_{-0.26}$ \\ $(68.65)$} & \makecell{$0.811^{+0.007}_{-0.008}$ \\ $(0.808)$} & --- & --- & 4248.0 & 53.8 \\
         & DQCD & \makecell{$0.02294^{+0.00014}_{-0.00013}$ \\ $(0.02297)$} & \makecell{$0.1279^{+0.0022}_{-0.0021}$ \\ $(0.1287)$} & \makecell{$3.056^{+0.012}_{-0.015}$ \\ $(3.052)$} & \makecell{$0.9729^{+0.0034}_{-0.0033}$ \\ $(0.9725)$} & \makecell{$0.0646^{+0.0062}_{-0.0086}$ \\ $(0.0611)$} & \makecell{$71.60^{+0.63}_{-0.63}$ \\ $(71.82)$} & \makecell{$0.817^{+0.011}_{-0.008}$ \\ $(0.818)$} & \makecell{$0.535^{+0.102}_{-0.098}$ \\ $(0.577)$} & \makecell{$6.11^{+0.39}_{-0.13}$ \\ $(6.04)$} & 4217.6 & 24.0 \\
        \hline\hline
    \end{tabular}}
    \caption{Bayesian posteriors for the cosmological parameters fit to the specified combinations of datasets. Our baseline dataset consists of the {\em Planck} CMB dataset \cite{Planck:2018vyg, 2022JCAP...09..039C}, DESI DR2 \cite{2025PhRvD.112h3515A}, and the Pantheon+ \cite{2022ApJ...938..110B} supernovae data. Each row displays the Bayesian posterior range for the listed parameter fit to either $\Lambda$CDM or the dark QCD model. Beneath that, we indicate the parameter value that maximizes the likelihood of the data. The amplitude of the matter density clustering $S_8$ is not a fundamental parameter, but is derived from the cosmology. For each dataset and model, $\Delta \chi^2$ is calculated relative to the $\chi^2$ of the baseline in that model, as per the conventions of \cite{Schoneberg:2026eys,Schoneberg:2026vaf}.}
    \label{tab:fits}
\end{table*}

In addition to the Hubble tension, the angular size of the sound horizon at the drag epoch $r_d$ measured by DESI through the BAO peak in large scale structure also presents some tension within $\Lambda$CDM. The ratio of distance measures ($D_M$, $D_V$, and $D_H$) to $r_d$ as seen in DESI data are $2.3\sigma$ discrepant with the cosmological parameters fit to the early Universe data \cite{2025PhRvD.112h3515A}. This has led to considerable observational and theoretical interest, and models of evolving dark energy are preferred over $\Lambda$CDM by $\sim 3\sigma$ \cite{DESI:2025fii,DESI:2025wyn}, though such models typically contain a ``phantom'' crossing where the equation of state $w$ drops below $-1$. 

\begin{figure}[t]
\centering
\includegraphics[width=0.9\textwidth]{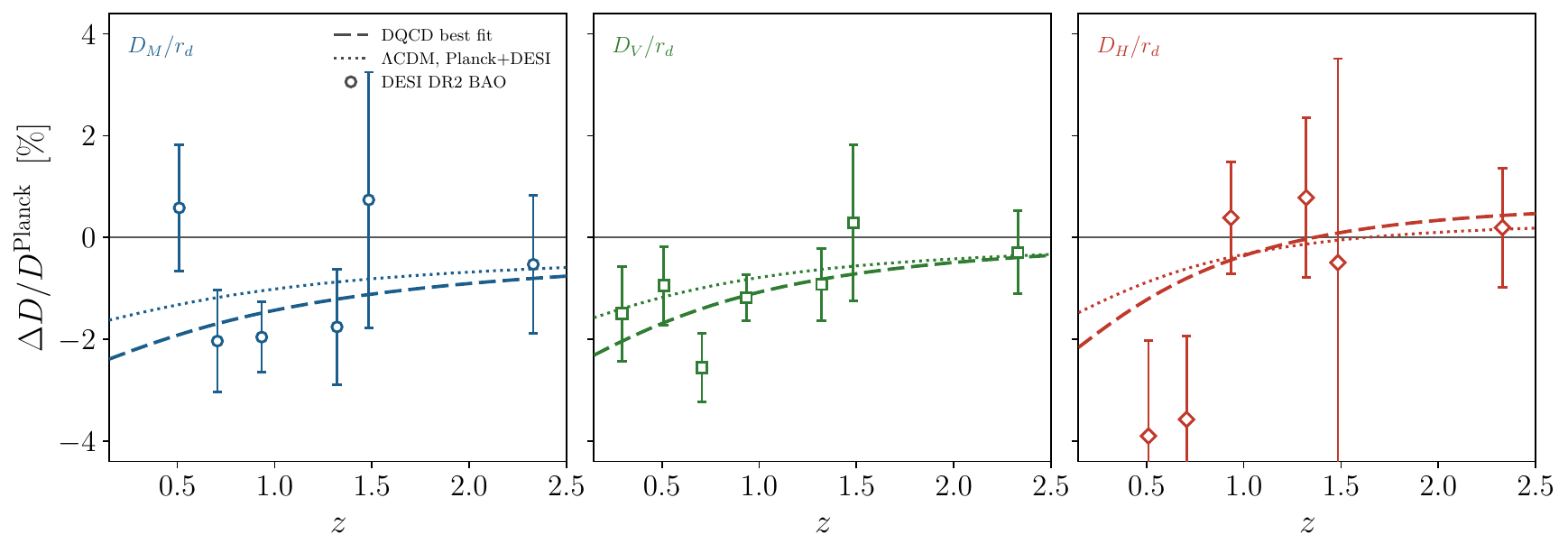}
\caption{The shift relative to the $\Lambda$CDM values assuming a cosmology fit to {\em Planck} \cite{Planck:2018vyg} of the ratios of the transverse moving distance $D_M$, the dilation distance $D_V$, and the Hubble distance $D_H$ to the sound horizon at the drag epoch $r_d$. DESI DR2 \cite{2025PhRvD.112h3515A} measurements are shown as data points and errors (note these three measurements are not independent). The predictions from $\Lambda$CDM using parameters fit to both {\em Planck} and DESI DR2 are shown as dotted lines, and the predictions of the dark QCD model using the maximum-likelihood parameters are shown as dashed lines. \label{fig:desi} }
\end{figure}

However, models of the early Universe that prefer different values of $\Omega_m$ also shift the distance measures at low $z$ relative to the {\em Planck} baseline, and so can resolve some of the tension as well. In Figure~\ref{fig:desi}, we plot the shift of the distance ratios $D_M/r_d$, $D_V/r_d$, and $D_H/r_d$ relative to the $\Lambda$CDM expectation using cosmological parameters derived only from {\em Planck} data, with the shift calculated for both a $\Lambda$CDM model with a joint fit to both {\em Planck} and DESI DR2, and for DQCD with parameters that provide to the maximum likelihood across all datasets. 
As can be seen, the dark QCD model parameters which provide the greatest reduction in the Hubble Tension significantly reduce the tension with DESI measurements relative to the {\em Planck}-only $\Lambda$CDM cosmology, and mildly outperform $\Lambda$CDM with parameters fit to both {\em Planck} and DESI.

\begin{figure}[t]
\centering
\includegraphics[width=0.9\textwidth]{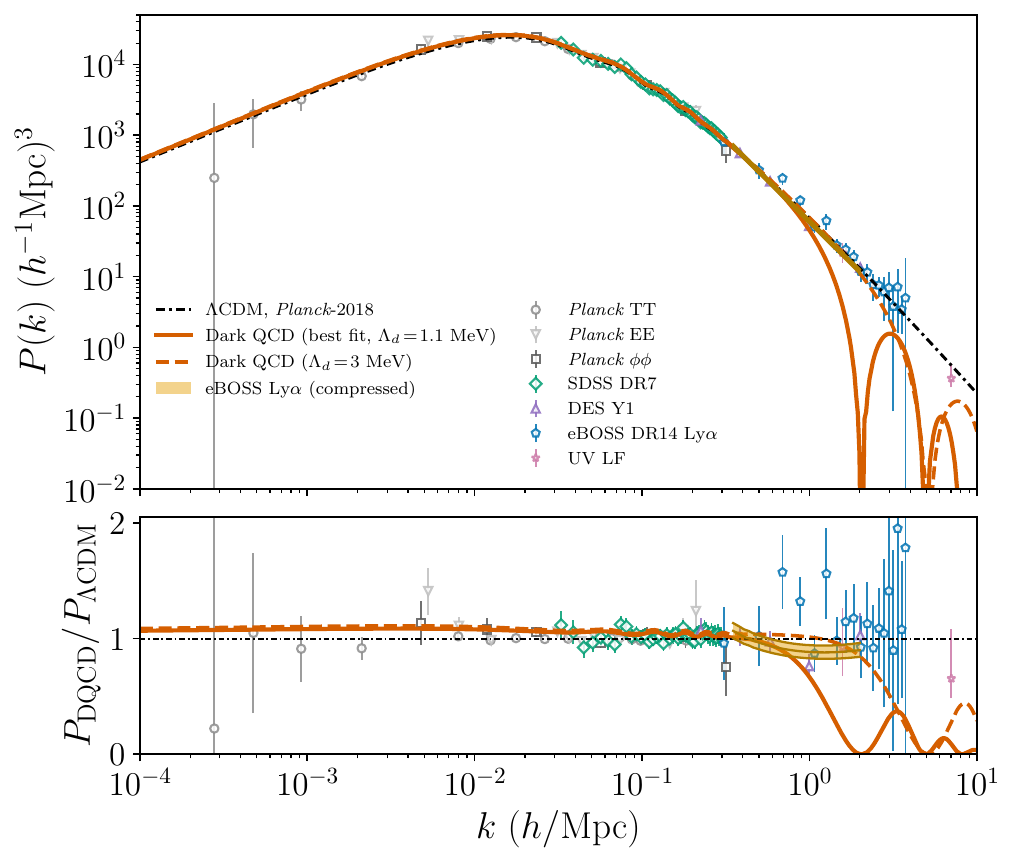}
\caption{The matter power spectrum as calculated by \textsc{Class} for $\Lambda$CDM using the {\em Planck} best-fit parameters (black, dot-dashed) and the dark QCD model with the maximum likelihood parameters (orange solid) and with $\Lambda_{\rm D} = 3$~MeV (orange dashed). We show the measurements of the power spectrum from {\rm Planck} $TT$ and $EE$ \cite{Planck:2019nip}, {\em Planck} lensing \cite{Planck:2018lbu}, SDSS DR7 \cite{2010MNRAS.404...60R}, DES \cite{DES:2017qwj}, eBoss DR14 \cite{eBOSS:2018qyj}, and {\em Hubble} UV LF \cite{Sabti:2021unj}. We show the compressed likelihood from eBoss Lyman-$\alpha$ data in yellow \cite{Pedersen:2020kaw,Goldstein:2023gnw}. The lower panel shows the ratio of the power spectrum to the {\em Planck} $\Lambda$CDM expectation. \label{fig:matter_power_spectrum} }
\end{figure}

Lastly, we consider the matter power spectrum, is modified through dark acoustic oscillations \cite{Cyr-Racine:2012tfp,Cyr-Racine:2013fsa,Cyr-Racine:2015ihg} mediated by the dark radiation at scales corresponding to the horizon at time of kinetic decoupling Eq.~\eqref{eq:zkdscaling}. In Figure~\ref{fig:matter_power_spectrum}, we show the matter power spectrum (as calculated by \textsc{Class}) both for $\Lambda$CDM using the {\em Planck} cosmology and the DQCD model using the maximum-likelihood parameters of Table~\ref{tab:fits}. We further show measurements of the power spectrum from {\em Planck} \cite{Planck:2019nip,Planck:2018lbu}, SDSS DR7 \cite{2010MNRAS.404...60R}, and DES \cite{DES:2017qwj}, as well as eBoss DR14 \cite{eBOSS:2018qyj} and {\em Hubble} UV LF \cite{Sabti:2021unj} Lyman-$\alpha$ measurements. Lyman-$\alpha$ constraints rely on simulation which assumes a particular cosmology. Applying them to new models which modify these assumptions has the potential for error. We include the compressed likelihood model fit to the SDSS eBoss data \cite{McDonald:1999dt,Pedersen:2019ieb,Pedersen:2020kaw,Goldstein:2023gnw} which has been shown to be valid across differing cosmologies. It should also be noted that the {\em Planck} $\Lambda$CDM cosmology is itself in some tension with the Lyman-$\alpha$ data \cite{Rogers:2023upm}, though this may be analysis dependent \cite{Fernandez:2023grg}.

As can be seen, the dark QCD model with the maximum likelihood parameters is in significant tension with the Lyman-$\alpha$ measurements. Similar tensions are seen in other solutions to the Hubble Tension \cite{Hill:2020osr,Ivanov:2020ril,Goldstein:2023gnw}, in particular to models of dark radiation with a dark matter drag rate \cite{Bagherian:2024obh}.

In the dark QCD model, the scale of these oscillations is set largely by $\Lambda_{\rm D}$ through $z_{\rm kd}$. When fit to cosmological data, we have found that the posterior is relatively flat for a large range of $\Lambda_{\rm D}$ above 1~MeV. All confinement scales in this region are compatible with high values of $H_0$ (see Figure~\ref{fig:likelihood_corner}).

We show as an example the matter power spectrum for a parameter point with $\Lambda_{\rm D} =3$~MeV and $H_0 = 72.23$~km/s/Mpc (with $Q_{\rm DMAP} = 3.3\sigma$). This parameter point overshoots the compressed likelihood, which is the same behavior as seen in $\Lambda$CDM. Larger $\Lambda_{\rm D} \sim 4-5$~MeV would have identical power spectra to $\Lambda$CDM through the entire $k$-range that Lyman-$\alpha$ is sensitive to, without significantly degrading the fit to SH0ES and CMB data.

\section{Low Scale Dark QCD in the Present Universe \label{sec:presentday}}

At first glance, a model in which TeV-scale dark matter interacts under a confining dark gauge group with $\Lambda_{\rm D} \approx $~MeV would seem to be excluded by limits on self-interacting cross sections for dark matter \cite{Spergel:1999mh,Ackerman:2008kmp,Kaplan:2009de,Cyr-Racine:2012tfp,Tulin:2013teo,Zavala_2013,Tulin:2017ara,Ghalsasi:2017jna}. In particular, the Bullet Cluster \cite{Clowe:2006eq,Randall:2008ppe} constrains the velocity-independent cross section to be $\sigma/m_\chi \lesssim 1~{\rm cm^2/g} \approx 4500~{\rm GeV}^{-3}$. Our model has a total self-interaction cross section of $\sim \pi \Lambda_{\rm D}^{-2}$, resulting in $\sigma/m_\chi \sim  3 \times 10^3~{\rm GeV}^{-3}$, at best only marginally allowed. However, this cross section is misleading, and a closer examination reveals the the model safely evades these limits while containing potentially interesting novel phenomenology in the late Universe.

The $\propto \Lambda_{\rm D}^{-2}$ cross section is the total ``black disk'' scattering cross section of two hadronized massive particles colliding via their DQCD interaction (the equivalent of neutron-neutron scattering). However, this cross section corresponds to all possibly scattering processes. The majority of these will be interactions between the ``muck'' of the light quarks and gluons surrounding the heavy quark. These interactions will -- for the most part -- only result in the transfer of ${\cal O}(\Lambda_{\rm D})$ momentum between the dark matter. Hard interactions that transfer significant momentum require the two heavy $Q$ quarks to interact; these particles are carrying $m_\chi v \gg \Lambda_{\rm D}$ momentum (assuming velocities characteristic of spiral galaxies or clusters). As a result, the interaction between the two heavy quarks can only occur at high energies or close ranges, when the gauge force is perturbative:
\begin{equation}
    \frac{\sigma}{m_\chi} \approx \frac{\alpha_{\rm D}^2}{4\pi m_\chi^3}.
\end{equation}
Here $\alpha_{\rm D}$ should be evaluated at the characteristic momentum of the system. This is well below the self-interacting constraints even for confinement scales much smaller those we found ease the Hubble Tension in Section~\ref{sec:data}. In principle, even $\Lambda_{\rm D}$ on the order of the eV scale could be compatible with observations of dark matter systems in the late Universe, due to the hierarchy of scales between $\Lambda_{\rm D}$ and $m_\chi$. In such low-scale dark QCD models, the halos of galaxies and galaxies clusters exist in an odd phase of matter: the interparticle spacing is larger than $\Lambda_{\rm D}^{-1}$ while the average scattering occurs at energy scales much larger than $\Lambda_{\rm D}$.

The major concern then is not energy exchange via direct dark matter-dark matter scattering, but rather energy loss through emission of light pions during scattering events and energy transfer through pion reabsorption. Each black disk scattering produces ${\cal O}(1)$ pions \cite{Weinberg:1970bs,Bjorken:1992rv,Shuvaev:1997cs} carrying away ${\cal O}(\Lambda_{\rm D})$ from the event. The cooling timescale for this process to carry away the majority of a dark matter particles kinetic energy is
\begin{equation}
    t_c \approx \frac{m_\chi v \Lambda_{\rm D}}{2\rho_\chi},
\end{equation}
whereas the free-fall time is
\begin{equation}
    t_f \approx \left(\frac{16}{3\pi} G_N \rho_\chi \right)^{1/2}.
\end{equation}
Systems with $t_f < t_c$ will demonstrate no significant difference in their internal structure of dark matter relative to a standard non-interacting particle. This sets a lower bound of
\begin{eqnarray}
    m_\chi^2 & > & \sqrt{\frac{3\pi}{4}\rho_\chi}\frac{M_{\rm Pl}}{\Lambda_{\rm D} v} \\
     & > & (150~{\rm GeV})^2\left(\frac{\rho_\chi}{0.3~{\rm GeV/cm^3}}\right)^{1/2}\left(\frac{{\rm MeV}}{\Lambda_{\rm D}} \right)\left(\frac{10^{-3}}{v} \right). \nonumber
\end{eqnarray}
For the benchmark parameters we work with in this paper, the cooling time is longer than the free-fall time and so results in no modification of galactic structure on the scale of spiral or dwarf galaxies. Note in particular however that our choice of $m_\chi \approx 1$~TeV is only weakly constrained by the early Universe data relative to $\Lambda_{\rm D}$, and smaller values are possible where non-trivial cooling can occur.

Even for the parameters of our benchmark model, there can be significant cooling in gravitationally bound structures smaller than dwarf galaxies (with lower characteristic velocities). Such cooling would significantly modify small-scale substructure within spiral galaxies without violating current observational constraints \cite{Fan:2013yva,Buckley:2017ttd,Buckley:2024eoe}.

Existing models of dark matter with effective cooling channels typically allow the dark radiation to escape high-density regions of dark matter, resulting in the gravitothermal collapse of the system when the loss-rate outpaces the free-fall time \cite{
Kaplan:2009de,Fan:2013bea,DAmico:2017lqj,Buckley:2017ttd,Buckley:2024eoe}. However, the dark QCD model can -- for appropriate choices of parameters -- result in low-opacity systems where the loss rate of energy through dark radiation is high and the mean free path is low. Such a configuration changes how a halo of dark matter will fragment during the cooling process, and can result in final seeds more massive than found in previously studied dissipative dark matter models. We consider the implications of this class of low-opacity dark matter models for galactic structure in future work.

\section{Conclusions \label{sec:conclusions}}

We have presented a new model of the dark sector containing a confining gauge group that can be analogized as ``dark QCD.'' Notably, we assume a hierarchy of scales such that the majority of the dark matter is composed of a heavy stable ``dark quark'' akin to a stable top quark. The stability occurs in our model as the dark sector has no equivalent of the weak interaction.
This particle is much heavier than the confinement scale, and our model also contains light ``dark quarks'' with masses far beneath the dark QCD confinement scale $\Lambda_{\rm D}$. This is the same rough hierarchy of fermion masses and confinement scale as exists within the Standard Model.

As the dark sector cools in the expanding Universe, the dark quarks confine. Dark matter today is thus a tower of hadronic states containing a single heavy dark quark, with mass splittings $<\Lambda_{\rm D}$. The dark radiation converts from a mix of dark gluons and light dark quarks to dark pions at confinement; as the light quarks have mass far below $\Lambda_{\rm D}$, the dark pions remain relativistic long after the phase transition. Critically, the dark radiation is highly viscous. Viscous dark radiation in the early Universe suppresses oscillations in the high-$\ell$ multipoles of the CMB relative to dark sectors with perfect fluids or free-streaming relativistic particles.

We find that this model of dark matter and dark radiation with a confinement scale around $1$~MeV can improve the global fit to cosmological data relative to standard $\Lambda$CDM. In particular, we can ease the Hubble tension between CMB and the late-time measurements to $\sim 3\sigma$, with a posterior mean Hubble parameter of $H_0 = 71.60\pm0.63~{\rm km/s/Mpc}$,\footnote{The most-likely value is $71.82~{\rm km/s/Mpc}$.} without increasing the tension with measurements of $S_8$. Our model also has improved fits to the DESI DR2 measures of the BAO angular scale relative to the standard $\Lambda$CDM parametrization. 
This level of agreement within datasets appears to be competitive with the ``Finalist'' models identified by the recent Hubble World Cup analysis \cite{Schoneberg:2026eys,Schoneberg:2026vaf} of proposed solutions to the Hubble tension.

The best-fit parameter point for the dark QCD model result in dark acoustic oscillations, and are in disagreement with Lyman-$\alpha$ measurements. Increasing the confinement scale from $\sim 1$~MeV to 3~MeV or higher would evade these constraints without significantly altering the fit to other cosmological data or creating a preference for a lower value of $H_0$. The existence of dark acoustic oscillations at scales close to current measurements motivates additional work in this area to produce robust limits.

All else being equal, increasing the Hubble parameter today can result in a prediction that the Universe is younger than the oldest objects within it. This consideration is relevant for many solutions to the Hubble tension \cite{Lin:2019htv,Jimenez:2019onw,Bernal:2021yli,Boylan-Kolchin:2021fvy}. The dark QCD model fit to all datasets (including SH0ES) gives an age of the Universe of $13.22\pm 0.10$~Gyr. This is younger than the standard $\Lambda$CDM value, but within $1\sigma$ of recent estimates of the Universe's age from globular clusters $13.5\pm 0.27$~Gyr \cite{Valcin:2020vav,Valcin:2021jcg}. Measurements of the ages of stellar populations place the age of the Universe as greater than 13~Gyr at 90\%CL \cite{Tomasetti:2025jpg}, compatible with the DQCD model.

In the present day, such strongly-interacting dark matter would seem to be constrained by measurements of dark matter self-interactions. However, while the total self-interaction cross section of the confined heavy dark hadrons is large, most of this cross section is the result of low-energy interactions between the DQCD ``muck'' surrounding the heavy dark quark. Significant momentum transfer between dark matter particles occurs only at small scales and high energies, when the interaction is perturbative. 

More relevant to the structure of dark matter halos today is repeated low-energy scattering resulting in the emission of dark pions. For the best-fit parameters for the early Universe, such energy loss mechanisms are not expected to modify the structure of dark matter halos at the scales where we have observational constraints. However, they could modify structure at smaller scales, and would have novel cooling behavior with lower opacity than other dissipative models of dark matter. The dark QCD model can therefore modify small scale structure in two ways: primordially through dark acoustic oscillations, and evolutionarily through cooling and heat transfer. 

In order to focus on the novel phenomenology within the dark sector that can relieve the tensions within cosmological datasets, the minimal model we consider in this work does not have any interactions with the Standard Model. Adding in Standard Model couplings can result in further interesting phenomenological signatures \cite{Kribs:2009fy}, due to the closely-spaced dark hadronic mass states and the low scale of $\Lambda_{\rm D}$. Combined with future measurements of cosmological parameters and small-scale structure within galaxies, such signals provide interesting pathways for further work.

\section*{Acknowledgments}

We thank Mitchell Weikert, Peizhi Du, and Ranit Das for discussions. The authors are supported by the DOE under Award Number DOE-SC0010008. Generative LLM output (Anthropic {\em Claude} v4 and v5) was used in the production of the modified \textsc{Class} code; no LLM output is used in the text of this paper.

\bibliography{darkQCD}

\end{document}